\documentclass[12pt]{iopart}

\usepackage{iopams}

\input epsf

\usepackage{graphicx}

\newcommand{\beq}{\begin{equation}}
\newcommand{\eeq}{\end{equation}}
\newcommand{\beqs}{\begin{eqnarray}}
\newcommand{\eeqs}{\end{eqnarray}}
\eqnobysec

\begin{document}

\title[Exact Potts/Tutte Polynomials for Polygon Chain Graphs]
{Exact Potts/Tutte Polynomials for Polygon Chain Graphs}

\author{Robert Shrock$^1$}

\address{$^1$ \ C. N. Yang Institute for Theoretical Physics,
Stony Brook University, Stony Brook, NY 11794}

\eads{\mailto{robert.shrock@stonybrook.edu}}

\begin{abstract}

We present exact calculations of Potts model partition functions and the
equivalent Tutte polynomials for polygon chain graphs with open and cyclic
boundary conditions. Special cases of the results that yield flow and
reliability polynomials are discussed. We also analyze special cases of the
Tutte polynomials that determine various quantities of graph-theoretic
interest.

\end{abstract}


%\keywords{} 

\maketitle

\newpage
\pagestyle{plain}
\pagenumbering{arabic}

\section{Introduction}
\label{intro}

In Ref. \cite{nec} with S.-H. Tsai, exact results were given for the partition
function of the zero-temperature $q$-state Potts antiferromagnet, or
equivalently, the chromatic polynomial, on open and cyclic chain graphs
composed of $m$ repetitions of $p$-sided polygons connected to each other by
line segments. In this paper we generalize this work and give the full Potts
model partition function for these families of graphs. The $q$-state Potts
model has long been of interest in the study of phase transitions and critical
phenomena \cite{wurev}.  On a lattice, or, more generally, on a graph $G$, at
temperature $T=1/(k_B\beta)$, the partition function for this model is $Z=
\sum_{\{\sigma_i\}}e^{-\beta {\cal H}}$, with the Hamiltonian ${\cal H} =
-J\sum_{e_{ij}}\delta_{\sigma_i \sigma_j}$, where $J$ is the spin-spin
interaction constant, $i$ and $j$ denote vertices on $G$, $e_{ij}$ is the edge
connecting them, and $\sigma_i$ are classical spins taking on values in the set
$\{1,...,q\}$. We use the notation $K=\beta J$ and $v=e^K-1$.  Thus, for the
Potts ferromagnet ($J > 0$) and antiferromagnet ($J < 0$), the physical ranges
of $v$ are $v \ge 0$ and $-1 \le v \le 0$, respectively.  For the Potts
antiferromagnet (PAF), $J < 0$ so that, as $T \to 0$, $K \to -\infty$; hence,
in this limit, the only contributions to the PAF partition function are from
spin configurations in which adjacent spins have different values.  The
resultant $T=0$ PAF partition function is therefore precisely the
chromatic polynomial $P(G,q)$ of the graph $G$, which counts the number of ways
of assigning $q$ colors to the vertices of $G$ subject to the condition that no
two adjacent vertices have the same color.  (These are called proper
$q$-colorings of $G$.)

\begin{figure}
\centering
\leavevmode
\epsfxsize=4.0in
\epsffile{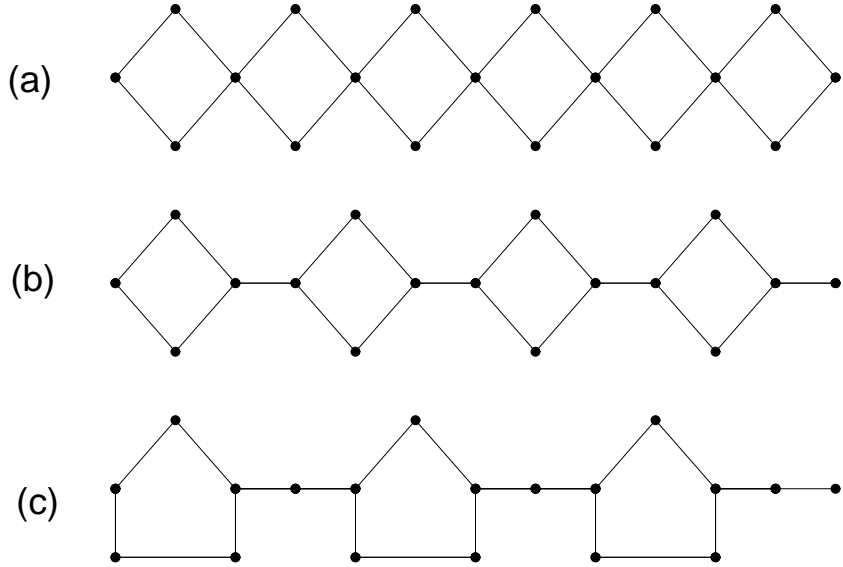}
\vspace{1 cm}
\caption{\footnotesize{Illustrations of cyclic and open polygon chain graphs
$G_{e_1,e_2,e_g,m}$ and $G_{e_1,e_2,e_g,m;o}$ with $(e_1,e_2,e_g,m)=$ (a)
(2,2,0,6), (b) (2,2,1,4), (c) (2,3,2,3).  For the cyclic (open) chain graphs, the
rightmost vertex on each graph is identified with (distinct from) the
leftmost vertex at the same level, respectively.}}
\label{pgchain}
\end{figure}

In general, a graph $G=(V,E)$ is defined by its set of vertices (sites), $V$,
and its set of edges (bonds), $E$.  We denote the number of vertices of $G$ as
$n=n(G)=|V|$ and the number of edges of $G$ as $e(G)=|E|$. The families of
graphs to be considered here are open and cyclic chains of polygons connected
by line segments. One may regard the chain as being oriented so that the
longitudinal direction is horizontal.  Each polygon is connected to the chain
at two vertices, such that there are $e_1$ edges of the polygon above the chain
and $e_2$ edges below the chain, and there are $e_g$ edges between each polygon
(where $g$ stands for gap).  Some illustrative examples are given in
Fig. \ref{pgchain} (from \cite{nec}).  The basic subgraph unit of the chain is
thus a polygon with
\beq
p = e_1+e_2
\label{poly}
\eeq
edges, connected to a line segment with $e_g$ edges.  The full chain with open
(o) or cyclic (c) boundary conditions (BC) is comprised of $m$ repetitions of
this basic subgraph comprised of the $p$-gon and $e_g$-length line segment, and
is denoted, as in \cite{nec}, by $G_{e_1,e_2,e_g,m;BC}$.  Since the two sides
of the chain are equivalent, the interchange $e_1 \leftrightarrow e_2$ leaves
it invariant, so $G_{e_1,e_2,e_g,m;BC} = G_{e_2,e_1,e_g,m;BC}$.  This implies
that all of the quantities to be presented below are also invariant under this
interchange.  Indeed, in some of these quantities, the numbers $e_1$ and $e_2$
only enter in the form of their sum, $p$. Since the $(m+1)$'th member of a
family of strip graphs of this sort can be obtained from the $m$'th member by
gluing on an additional basic subgraph unit or, in the case of the cyclic
strip, by cutting the strip transversely, inserting an additional basic
subgraph unit and regluing, these are recursive families, in the sense of
\cite{bds}.  The numbers of vertices and edges of these graphs are
\beq
n(G_{e_1,e_2,e_g,m;o}) = (p+e_g-1)m + 1 \ , 
\label{nonec}
\eeq
\beq
n(G_{e_1,e_2,e_g,m;c}) = (p+e_g-1)m \ , 
\label{nnec}
\eeq
and
\beq
e(G_{e_1,e_2,e_g,m;o}) = e(G_{e_1,e_2,e_g,m;c}) = (p+e_g)m  \ . 
\label{enec}  
\eeq
Clearly, $G_{e_1,e_2,e_g,m;o}$ and $G_{e_1,e_2,e_g,m;c}$ are planar graphs.  In
the context of statistical mechanics, one takes $e_1 \ge 1$ and $e_2 \ge 1$,
whence $p \ge 2$, since setting either $e_1=0$ or $e_2=0$ would mean that a
spin $\sigma_i$ would interact with itself rather than with neighboring
spins. However, in the context of mathematical graph theory, one may formally
consider the case where $e_1=0, \ e_2=1$ or $e_1=1, \ e_2=0$, whence, $p=1$.
In these cases, the graphs $G_{e_1,e_2,e_g,m;BC}$ contain loops, where a loop
is defined as an edge that connects a vertex back to itself.

One motivation for the present work is to understand how the results of
\cite{nec} can be generalized to finite temperature and ferromagnetic as well
as antiferromagnetic spin-spin couplings.  Another is to get further insight
into how properties of a graph affect the Potts partition function $Z$ or
equivalent nTutte polynomial (see further below).  A particular appeal of the
chain graphs considered here is that the results are sufficiently simple that
one can study them in considerable explicit detail.  For recursive strip graphs
of length $m$ basic subunits, $Z$ is a sum of $m$'th powers of certain
algebraic functions, generically denoted as $\lambda$'s. Although calculations
of $Z$ have been done on wider strips of regular lattices, as the strip width
increases, the results rapidly become quite complicated. For example, for the
cyclic (or M\"obius) strip graph of the square lattice of width $L_y=3$ and
length $L_x=m$, there are ${6 \choose 3}=20$ different $\lambda$'s, and many of
these are solutions of algebraic equations of sufficiently high degree so that
they cannot be expressed in closed analytic form \cite{s3a}.  Although for a
given width, fewer $\lambda$'s occur for an open strip than for a cyclic strip,
even for width $L_y=3$, there are five of these, including four that are
solutions of a 4'th order algebraic equation, rendering an explicit expression
rather cumbersome \cite{s3a,ts}.  Thus, it is valuable to investigate the
effects of graphical properties on $Z$ for families of strip graphs where one
can obtain explicit exact closed-form analytic solutions for the $\lambda$'s
that enter.

\section{General Background}

In this section we discuss some general background material relevant to our
study.  Let $G^\prime=(V,E^\prime)$ be a spanning subgraph of $G$, i.e. a
subgraph having the same vertex set $V$ and an edge set $E^\prime \subseteq
E$. Then $Z$ can be written as \cite{fk}
\beq
Z(G,q,v) = \sum_{G^\prime \subseteq G} q^{k(G^\prime)}v^{e(G^\prime)} \ , 
\label{cluster}
\eeq
where $k(G')$ denotes the number of connected components of $G'$.  As is
evident from (\ref{cluster}), $Z(G,q,v)$ is a polynomial in $q$ and $v$.  In
the ferromagnetic case with $v > 0$, Eq. (\ref{cluster}) allows one to extend
the definition of $q$ from the positive integers to the positive real numbers
while maintaining $Z(G,q,v) > 0$ and hence a Gibbs measure.  Since $k(G') \ge 1
\ \forall \ G'$, it follows that $Z(G,q,v)$ always has a factor of $q$. It is
thus convenient to define the reduced partition function $Z_r(G,q,v)$ as
\beq
Z(G,q,v) = q Z_r(G,q,v)
\label{zr}
\eeq
For a graph $G$, let us denote $G-e$ as the graph obtained by deleting the edge
$e$ and $G/e$ as the graph obtained by deleting the edge $e$ and identifying
the two vertices that were connected by this edge of $G$.  This operation is
called a contraction of $G$ on $e$. From Eq. (\ref{cluster}), it follows that
$Z(G,q,v)$ satisfies the deletion-contraction relation
\beq
Z(G,q,v) = Z(G-e,q,v)+vZ(G/e,q,v) \ . 
\label{dcr}
\eeq

The Potts model partition function is equivalent to the Tutte polynomial, 
$T(G,x,y)$, an object of considerable interest in mathematical graph theory.
For a graph $G$ \cite{bbook,boll}, 
\beq
T(G,x,y) = \sum_{G' \subseteq G} (x-1)^{k(G')-k(G)}(y-1)^{c(G')} \ , 
\label{t}
\eeq
where $c(G')$ denotes the number of (linearly independent) cycles in $G'$.
Note that $c(G)=e(G)+k(G)-n(G)$.  As is clear from (\ref{t}), $T(G,x,y)$ is a
polynomial in $x$ and $y$.  All of the families of graphs considered here are
connected, so that $k(G)=1$.  Let us define
\beq
x = 1 + \frac{q}{v} \ , \quad y = v+1  \ , 
\label{xyqv}
\eeq
so that $q=(x-1)(y-1)$. Then the equivalence between $Z(G,q,v)$ and $T(G,x,y)$
is given by 
\beq
Z(G,q,v) = (x-1)^{k(G)}(y-1)^{n(G)}T(G,x,y) \ . 
\label{ztrel}
\eeq
The special case $v=-1$ defines the $T=0$ Potts antiferromagnet, and sets the
variables in the Tutte polynomial equal to $x=1-q$ and $y=0$; in this case 
these functions yield the chromatic polynomial: 
\beq
Z(G,q,-1) = P(G,q) = (-q)^{k(G)}(-1)^{n(G)}T(G,1-q,0) \ . 
\label{chrompoly}
\eeq
If $G$ contains a loop, then $T(G,x,y)$ contains a factor of $y$ and $P(G,q)$
vanishes identically, since it is not possible to satisfy the proper
$q$-coloring condition.

We recall two elementary results.  For any tree graph $T_n$ with $n$ vertices, 
\beq
T(T_n,x,y)=x^{n-1} , \quad Z(T_n,q,v) = q(q+v)^{n-1} \ . 
\label{tztn}
\eeq
For the circuit graph with $n$ vertices, $C_n$, 
\beq
T(C_n,x,y) = \frac{x^n + c_1}{x-1} \ , \quad 
Z(C_n,q,v) = (q+v)^n + (q-1)v^n \ . 
\label{tzcn}
\eeq
where $c_1=xy-x-y=q-1$. Note that although $T(C_n,x,y)$ is expressed in
Eq. (\ref{tzcn}) as a rational function, it is actually a polynomial, as is
guaranteed by Eq. (\ref{t}).  Explicitly, $T(C_1,x,y)=y$ and, for $n \ge 2$,
$T(C_n,x,y)=y+\sum_{j=1}^{n-1} x^j$.  A similar comment applies to expressions
below involving $(x-1)$ denominators.  For the special case $v=-1$, since
$P(C_n,q)$ has a factor $q(q-1)$ for $n \ge 2$, it is convenient to define the
polynomial $D_n$ via
\beq
D_n = \frac{P(C_n,q)}{q(q-1)} = \sum_{s=0}^{n-2}(-1)^s {n-1 \choose s} 
q^{n-2-s} \ . 
\label{dk}
\eeq
(For $n=1$, $P(C_1,q)=0$, since $C_1$ is a vertex with a loop; hence also 
$D_1 = 0$.) 

\section{Calculations and Results}

Using a systematic application of the deletion-contraction theorem, we have
calculated the Potts partition function and equivalent Tutte polynomial for the
open and cyclic chain graphs $G_{e_1,e_2,e_g,m;o}$ and $G_{e_1,e_2,e_g,m;c}$.
For the open chain graph $G_{e_1,e_2,e_g,m;o}$ we find, for the Tutte
polynomial, 
\beq
T(G_{e_1,e_2,e_g,m;o},x,y) = (\lambda_{T,0})^m \ , 
\label{tonec}
\eeq
where
\beq
\lambda_{T,0} = T(T_{e_g+1},x,y) \, T(C_p,x,y) = 
x^{e_g} \Big ( \frac{x^p+c_1}{x-1} \Big ) \ . 
\label{tlamd0} 
\eeq
Note that $\lambda_{T,0}$ only depends on $e_1$ and $e_2$ via their sum, $p$. 
In general, if a graph $G$ can be expressed as $G_1 \cup G_2$ such that $G_1
\cap G_2$ is a single vertex, then $T(G,x,y)=T(G_1,x,y)T(G_2,x,y)$.  This
factorization property holds for the basic subgraph units of
$G_{e_1,e_2,e_g,m;o}$, since the $p$-gon intersects the line segment at a
single vertex.  The factorization property also holds, {\it a fortiori}, for
the open chain $G_{e_1,e_2,e_g,m;o}$ itself.  These facts imply the properties 
that (i) $\lambda_{T,0}$ has the form of a product of the Tutte polynomials of 
the tree graph $T_{e_g+1}$ and of the circuit graph, $C_p$, and (ii) 
$T(G_{e_1,e_2,e_g,m;o},x,y)$ has the form of a power of a single polynomial, as
given in Eq. (\ref{tonec}). Corresponding comments apply to
$Z(G_{e_1,e_2,e_g,m;o},q,v)$. 

For the cyclic chain graph $G_{e_1,e_2,e_g,m;c}$ we calculate 
\beq
T(G_{e_1,e_2,e_g,m;c},x,y) = \frac{1}{x-1}\bigg [ (\lambda_{T,0})^m + 
c_1(\lambda_{T,1})^m \bigg ] \ , 
\label{tnec}
\eeq
where $\lambda_{T,0}$ and $c_1$ were given above and 
\beq
\lambda_{T,1} = T(C_{e_1},x,y) + T(C_{e_2},x,y) + 1-y \ . 
\label{tlamd1}
\eeq
It is noteworthy that $\lambda_{T,1}$ is independent of $e_g$ and only depends
on the properties of the polygons, as encoded in their edge numbers $e_1$ and
$e_2$. 

From these general results, one can consider special cases of the various
graphical edge numbers $e_1$, $e_2$, and $e_g$.  For example, for the open and
cyclic chain graphs $G_{2,2,e_g,m;o}$ and 
$G_{2,2,e_g,m;c}$, illustrated (for $e_g=0,1$ and $m=6,4$) in Fig. 
\ref{pgchain}, we have 
\beq
T(G_{2,2,e_g,m;o},x,y) = [x^{e_g}(x+x^2+x^3+y)]^m
\label{tonec22egm}
\eeq
and
\beqs
T(G_{2,2,e_g,m;c},x,y) & = & \frac{1}{x-1} \bigg [ \{x^{e_g}(x+x^2+x^3+y)\}^m 
\cr\cr
& + & (xy-x-y)(2x+y+1)^m \bigg ] \ . 
\label{tnec22egm}
\eeqs

Various evaluations of $T(G_{e_1,e_2,e_g,m;o},x,y)$ and
$T(G_{e_1,e_2,e_g,m;c},x,y)$ for special values of the arguments $x$ and $y$
are of interest. In particular, for $x=1$, we have
\beq
T(G_{e_1,e_2,e_g,m;o},1,y) = (p+y-1)^m
\label{tonec_xeq1}
\eeq
and
\beqs
T(G_{e_1,e_2,e_g,m;c},1,y) & = & (p+y-1)^{m-1}\Bigg [ (me_g+y-1)(p+y-1) \cr\cr
& + &  \frac{m}{2}\bigg \{ p(p-1) - e_1(e_1-1) - e_2(e_2-1) \bigg \} \Bigg ]
 \ . 
\label{tnec_xeq1}
\eeqs

\section{Potts Model Partition Functions} 

The equivalent Potts model partition functions are
\beq
Z(G_{e_1,e_2,e_g,m;o},q,v) = q(\lambda_{Z,0})^m
\label{zonec}
\eeq
and
\beq
Z(G_{e_1,e_2,e_g,m;c},q,v) = (\lambda_{Z,0})^m + (q-1)(\lambda_{Z,1})^m \ , 
\label{znec}
\eeq
where
\beqs
& & \lambda_{Z,0} = Z_r(T_{e_g+1},q,v) \, Z_r(C_p,q,v) = (q+v)^{e_g}q^{-1} 
\Big [ (q+v)^p+(q-1)v^p \Big ] \cr\cr
& & 
\label{zlamd0} 
\eeqs
and 
\beqs
& & \lambda_{Z,1} = v^{e_g}\bigg [ v^{e_2}Z_r(C_{e_1},q,v)+
                                   v^{e_1}Z_r(C_{e_2},q,v) - v^p \bigg ]
\cr\cr
&=&v^{e_g}\Bigg [ q^{-1} \bigg [ 
v^{e_2}\Big \{ (q+v)^{e_1} +(q-1)v^{e_1} \Big \} + 
v^{e_1}\Big \{ (q+v)^{e_2} +(q-1)v^{e_2} \Big \} \bigg ] - v^p \Bigg ] \ . 
\cr\cr
& & 
\label{zlamd1}
\eeqs
We note that 
\beq
\lambda_{Z,0} = \lambda_{Z,1} = v^{p+e_g-1}(p+v) \quad {\rm at} \ q=0 
\label{lamq0}
\eeq
and
\beq
\lambda_{Z,1} = v^{e_g+1}\Big [ 2(v+1)-v^{p-1} \Big ] \quad {\rm for} \quad
e_1=e_2=1 \ . 
\label{zlamd1_e1e2equal1}
\eeq

\section{Chromatic Polynomials}

For the special case $v=-1$, our results for $Z(G_{e_1,e_2,e_g,m;o},q,v)$ and
$Z(G_{e_1,e_2,e_g,m;c},q,v)$ reduce to to the chromatic polynomials given 
(for $e_1 \ge 1$ and $e_2 \ge 1$) in \cite{nec}, 
\beq
P(G_{e_1,e_2,e_g,m;o},q) = q(\lambda_{P,0})^m
\label{ponec}
\eeq
and
\beq
P(G_{e_1,e_2,e_g,m;c},q) = (\lambda_{P,0})^m+(q-1)(\lambda_{P,1})^m \ , 
\label{pnec}
\eeq
where 
\beq
\lambda_{P,0} = (q-1)^{e_g+1}D_p
\label{lamp0}
\eeq
and
\beqs
& & \lambda_{P,1} = (-1)^{p+e_g} \, q^{-1} \, \Big [ (1-q)^{e_1}+(1-q)^{e_2}+
q-2 \Big ] \cr\cr
& = &  (-1)^{p+e_g}\biggl [ 1-p - \sum_{s=2}^{e_1}{e_1 \choose s}(-q)^{s-1}
- \sum_{s=2}^{e_2}{e_2 \choose s}(-q)^{s-1} \biggr ] \ . 
\label{lamp1}
\eeqs
(If either $e_1=0$ or $e_2=0$, the graphs contain one or more loops and the
chromatic polynomial vanishes identically.)

\section{Free Energy} 

It is of interest to remark on some thermodynamic properties of the $q$-state
Potts model on the infinite-length limit of these polygon chain graphs.
Although our results for the partition function apply for general $q$, we shall
restrict our attention here to integer $q \ge 2$. We denote the $m \to \infty$
limit of the chain graph $G_{e_1,e_2,e_g,m,BC}$ as $\{G_{e_1,e_2,e_g;BC}\}$.
The reduced, dimensionless free energy per vertex for the Potts model in this
limit is independent of boundary conditions, so we drop the $BC$ subscript.  We
find
\beqs
f(\{G_{e_1,e_2,e_g}\},q,v) & = & \lim_{m \to \infty} \ \frac{1}{n} 
\ln \Big [ Z(G_{e_1,e_2,e_g,m;BC}) \Big ] \cr\cr
& = & \frac{1}{(p+e_g-1)} \, \ln (\lambda_{Z0}) \cr\cr
& = & \frac{1}{(p+e_g-1)} \, \ln \Big [ (q+v)^{e_g} \, \{(q+v)^p + (q-1)v^p \} 
\ \Big ] \ . \cr\cr
& & 
\label{fonec}
\eeqs
(The actual Gibbs free energy per vertex is ${\cal G}=-k_BT f$.)  As was
discussed in \cite{a}, if one were to start with a Potts model partition
function $Z(G,q,v)$ for a cyclic strip graph $G$ with variable real (positive)
$q$, then, for a set of special (s) values of $q$, denoted $\{q_s\}$, one could
encounter the noncommutativity
\beq
\lim_{n \to \infty} \lim_{q \to q_s} Z(G,q,v)^{1/n} \ne
\lim_{q \to q_s} \lim_{n \to \infty} Z(G,q,v)^{1/n} \ .
\label{fnoncomm}
\eeq
For $G_{e_1,e_2,e_g,m;c}$, $\{q_s\}=\{0,1\}$. Since we are only interested in
(integral) $q \ge 2$ here, we do not encounter any such noncommutativity.

Because these chain graphs are quasi-one-dimensional, the Potts model (with
either sign of $J$) does not exhibit any critical behavior at nonzero
temperature.  However, it is of interest to investigate how thermodynamic
quantities depend on the graphical parameters $e_1$, $e_2$, and $e_g$.  A first
observation is that the reduced free energy $f$ and hence quantities that are
obtained as derivatives of $f$ with respect to temperature $T$, such as the
internal energy per site, $U=-\partial f/\partial \beta$ and specific heat per
site, $C = dU/dT$, only depend on $e_1$ and $e_2$ through their sum, $p$.  This
property follows from the fact that (in the limit $m \to \infty$) $f$ is
determined completely by the dominant $\lambda$ (i.e., the $\lambda$ with the
largest magnitude), namely $\lambda_{Z,0}$, and this only depends on $e_1$ and
$e_2$ via their sum, $p$.  As an explicit example, the internal energy per site
is
\beq
U =  -\frac{J(v+1)}{(p+e_g-1)} \Bigg [ \frac{e_g}{q+v} + 
p \bigg [ \frac{(q+v)^{p-1}+(q-1)v^{p-1}}{(q+v)^p + (q-1)v^p} \bigg ] \ \Bigg ]
\ . 
\label{u}
\eeq
In comparison, we recall that for the Potts model on the line,
$U_{1D}=-J(v+1)/(q+v)$.  In the expression (\ref{u}) for $U$, one thus sees the
interplay of the underlying circuit graph including the $m$ line segments, each
of $e_g$ edges, with the $m$ $p$-sided polygons, in the limit as $m \to
\infty$.  The limit of Eq. (\ref{u}) as the temperature $T \to \infty$, i.e.,
$v \to 0$, is
\beq
U = - \frac{J}{q} \, \bigg ( \frac{p+e_g}{p+e_g-1} \bigg ) \quad {\rm at} \ v=0
\ . 
\label{uv0}
\eeq
Since $(p+e_g)/(p+e_g-1) > 1$, $U(v=0)$ is more negative than the analogous
infinite-temperature limit of $U$ for the Potts model on the line, namely
$U(0)_{1D} = -J/q$.  One can also consider the zero-temperature limit.  For the
antiferromagnet, this reverts back to the analysis of the chromatic polynomial,
as in \cite{nec}.  For the ferromagnet, as $\beta \to \infty$,
\beq
\lim_{\beta \to \infty} \, U = -J \, 
\bigg ( \frac{p+e_g}{p+e_g-1} \bigg ) \quad {\rm for} \ J > 0 \ . 
\label{ufmt0}
\eeq
Again, this is more negative than the corresponding expression for the Potts
ferromagnet on the line, which is $\lim_{\beta \to \infty} U_{1D} = -J$. In
both the high- and low-temperature limits, these differences can be attributed
to the additional spin-spin interactions due to the combination of the repeated
polygons attached to the underlying global circuit graph.

Finally, we note some limiting cases:
\beq
\lim_{e_g \to \infty} \/ f(\{G_{e_1,e_2,e_g}\},q,v) = \ln (q+v) \quad 
{\rm for \ fixed \ finite} \ p 
\label{f_eg_infty}
\eeq
and
\beq
\lim_{p \to \infty} \/ f(\{G_{e_1,e_2,e_g}\},q,v) = \ln (q+v) \quad 
{\rm for \ fixed \ finite} \ e_g \ . 
\label{f_p_infty}
\eeq
In both of these cases, the dimensionless free energy per site thus reduces to
that of the $n \to \infty$ limit of the Potts model on a line, $f=\ln(q+v)$.
This reduction can be ascribed to the dominance of one of the two structural
parts of the graph - the global circuit in Eq. (\ref{f_eg_infty}) and the
polygons in Eq. (\ref{f_p_infty}).

\section{Zeros of the Partition Function and Locus ${\cal B}$}

It is also of interest to consider the locus of zeros of the partition
function.  In the limit $m \to \infty$, some zeros may merge to form curves
denoted as the locus ${\cal B}$. It is convenient to consider this locus in the
$q$ plane for fixed temperature variable $v$. For the case $v=-1$, i.e., the
$T=0$ Potts antiferromagnet, ${\cal B}$ was studied in \cite{nec}. Here we can
use our general results to analyze the case of the Potts antiferromagnet at
temperatures above $T=0$ and the Potts ferromagnet. Thus, we consider ${\cal
B}$ for $v$ in the full physical range, $v \ge -1$.

For the open polygon chain graph, since $Z(G_{e_1,e_2,e_g,m;o},q,v)$ involves
just the $m$'th power of a single $\lambda$ term, it has a fixed set of
discrete zeros, independent of $m$, with no continuous locus, i.e., ${\cal B}
= \emptyset$.  For the $m \to \infty$ limit of the cyclic polygon chain graph,
${\cal B}$ is a nontrivial locus, determined by the equation
$|\lambda_{Z,0}|=|\lambda_{Z,1}|$. For this limit, with $v=-1$, several
properties of ${\cal B}$ were derived in \cite{nec}, namely, that ${\cal B}$
(i) is compact, (ii) passes through $q=0$, and (iii) encloses regions in the
$q$ plane.  With the notation $e_s$ and $e_\ell$ denoting the
\underline{s}maller and \underline{l}arger of $e_1$ and $e_2$, it was shown in
\cite{nec} that (iv) if $e_s=1$, then ${\cal B}$ is the circle $|q-1|=1$,
independent of the values of $e_\ell$ and $e_g$, so that $q_c=2$, where $q_c$
is defined as the maximal point at which ${\cal B}$ crosses the real axis
(which it always does for the $m \to \infty$ limit of these cyclic polygon
chain graphs).  Two additional properties concerning $q_c$ values for $v=-1$
(denoted (B5) and (B6) in \cite{nec}) were also shown.

We use the notation $\lim_{m \to \infty} G_{e_1,e_2,e_g,m;c} \equiv \{
G_{e_1,e_2,e_g;c} \}$.  For this $m \to \infty$ limit of the cyclic polygon
chain graph, we find that properties (i)-(iii) continue to hold.  The proof of
the compactness property (i) is a generalization of the proof for $v=-1$ given
in \cite{nec}; one uses the fact that a necessary and sufficient condition that
${\cal B}$ is noncompact in the $q$ plane, passing through $1/q=0$, is that the
equation $|\lambda_{Z,0}|=|\lambda_{Z,1}|$ has a solution for $1/q=0$.  To show
that this is not the case, we extract a factor of $q^{p+e_g-1}$ from both sides
of this equation and define $\lambda_{Z,d} \equiv q^{p+e_g-1}\bar\lambda_{Z,d}$
for $d=0,1$.  Dividing both sides by the factor $|q^{p+e_g-1}|$ we have
$|\bar\lambda_{Z,0}|=|\bar\lambda_{Z,1}|$.  Taking $1/q \to 0$, we see that
this equation cannot be satisfied because $\bar\lambda_{Z,0} \to 1$ while
$\bar\lambda_{Z,1} \to 0$. This proves the compactness of ${\cal B}$ for
general $v$.  Property (ii) follows because, as Eq. (\ref{lamq0}) shows,
$|\lambda_{Z,0}|=|\lambda_{Z,1}|$ at $q=0$. To show property (iii), we evaluate
$f$ in Eq. (\ref{fonec}) (away from $q_s$ values so as to avoid the
noncommutativity).  The result is that in region $R_1$, which contains the
semi-infinite real intervals $q > q_c$ and $q < 0$, $\lambda_{Z,0}$ has a
larger magnitude than $\lambda_{Z,1}$, so $f=(p+e_g-1)^{-1}\ln \lambda_{Z,0}$,
as in Eq. (\ref{fonec}), while in a region $R_2$ containing an interval of
small positive $q$ neighboring the origin, $\lambda_{Z,1}$ has a larger
magnitude than $\lambda_{Z,0}$, so $f=(p+e_g-1)^{-1}\ln \lambda_{Z,1}$.  If
these regions were not completely separated by the nonanalytic boundary ${\cal
B}$, then one could analytically continue $f$ from one to the other, but this
would lead to a contradiction, since $f$ has a different functional form in
these two regions.  This proves that ${\cal B}$ separates the $q$ plane into
regions. Property (iv) for $v=-1$ was a special consequence of the complete
intersection theorem for chromatic polynomials, namely that if a graph $G$ is
the union $G=G_1 \cup G_2$ such that the intersection $G_1 \cap G_2 = K_r$,
where $K_r$ is the complete graph on $r$ vertices (i.e, the graph in which each
vertex is connected to every other vertex by an edge), then $P(G,q) =
P(G_1,q)P(G_2,q)/P(K_r,q)$, where $P(K_r,q) = \prod_{s=0}^{r-1} (q-s)$.  This
theorem does not apply for $v \ne -1$, so the property (iv) for $v=-1$ no
longer holds for $v > -1$.

We have obtained a number of results on $q_c$ for the general case $v >
-1$.  As we have discussed above, for $\{ G_{e_1,e_2,e_g;c} \}$, 
the locus ${\cal B}$ crosses the real $q$ axis at $q=0$ and at a maximal point
denoted $q_c$.  (Depending on the values of $e_1$, $e_2$, $e_g$, and $v$, 
${\cal B}$ may cross the $q$ axis at additional points besides $q=0$ and
$q_c$.) Some exact results on $q_c$ are 
\beq
q_c(\{G_{1,1,0;c} \}) = -2v(v+2) 
\label{qc_110}
\eeq
\beq
q_c(\{G_{1,1,1;c} \}) = -v(v+3) 
\label{qc_111}
\eeq
and
\beq
q_c(\{G_{2,2,e_g;c} \}) = -2v \quad \forall \ e_g \ . 
\label{qc_22eg}
\eeq
For most cases, $q_c$ is a root of a higher-order equation.  For example, for
$\{G_{1,1,2;c} \}$, $q_c$ is the maximal real root of the cubic equation
\beq
q^3+v(v+4)q^2+v^2(2v+5)q+2v^3(v+2)=0 \ . 
\label{qc112}
\eeq
These calculations may be compared with our previous results for $q_c$ for the
infinite-length limits of some other strips with periodic longitudinal boundary
conditions.  Denoting $\lim_{n \to \infty} C_n$ as $\{C\}$, one has 
$q_c(\{C\}) = -2v$, the same as $q_c(\{G_{2,2,e_g;c}\})$. For the
cyclic or M\"obius square-lattice ladder strip \cite{a}, 
\beq
q_c( \{ sq \ lad\}) = -v(v+3) \ ,
\label{qcsqladder}
\eeq
the same as $q_c(\{G_{1,1,1;c} \})$. For the cyclic or M\"obius ladder strip of
the triangular lattice \cite{ta}, and also the self-dual (sd) square-lattice 
ladder strip, 
\beq
q_c( \{ tri \ lad \}) = q_c( \{sq_{sd} \ lad \}) = -\frac{v(2v+5)}{v+2} \ . 
\label{qcsqdualladder}
\eeq
These results for $q_c(\{G\})$ have the general property that, as $v$ increases
from $-1$ to 0, i.e., as the temperature increases from 0 to infinity for the
Potts antiferromagnet, they decrease monotonically from their $v=-1$ values to
0, and in the ferromagnetic region, $v \ge 0$, they are negative. One may also
consider the accumulation locus of partition function zeros in the $v$ plane
for fixed $q$.  However, this is of somewhat less interest than ${\cal B}$ in
the $q$ plane for fixed $v$, since these polygon chain graphs are
quasi-one-dimensional and hence (for $q \ge 2$) ${\cal B}$ does not cross the
real $v$ axis at any point corresponding to nonzero temperature.

\section{Flow Polynomials} 

An interesting problem in graph theory is the task of enumerating discretized
flows on the edges of a (connected) $G$ that satisfy flow conservation at each
vertex, i.e. for which there are no sources or sinks.  The flow on each edge
can take on any of $q$ values modulo $q$, (so $q=0 {\rm mod} q$).  One
arbitrarily chooses a direction for each edge of $G$ and assigns a discretized
flow value to it. The value zero is excluded, since it is equivalent to the
edge being absent from $G$; henceforth, we take a $q$-flow to mean implicitly a
nowhere-zero $q$-flow.  The flow or current conservation condition is that the
flows into any vertex must be equal, mod $q$, to the flows outward from this
vertex.  These are called $q$-flows on $G$, and the number of these is given by
the flow polynomial $F(G,q)$.  This is a special case of the Tutte polynomial
for $x=0$ and $y=1-q$:
\beq
F(G,q) = (-1)^{c(G)} \, T(G,0,1-q) \ . 
\label{flowpoly}
\eeq
A bridge on a graph $G$ is defined as an edge with the property that if it is
deleted, this increases the number of connected components of $G$ by one.  If a
(connected) graph $G$ contains any bridge, then the flow polynomial vanishes
identically. 
The open strip graph $G_{e_1,e_2,e_g,m;o}$ contains at least one
bridge if $e_g \ge 1$, and therefore it does not allow any $q$-flows: 
\beq
F(G_{e_1,e_2,e_g,m;o},q) = 0 \quad {\rm if} \ e_g \ge 1 \ . 
\label{flow_onec_egnonzero}
\eeq
For $e_g=0$, we find 
\beq
F(G_{e_1,e_2,e_g,m;o},q) = (q-1)^m \quad {\rm if} \ e_g = 0 \ . 
\label{flow_onec_egzero}
\eeq
Note that this is equal to $[F(C_p,q)]^m$, showing that the flows occur
independently in each $p$-gon circuit graph.  Algebraically, this follows from
the factorized form of $T(G_{e_1,e_2,e_g,m;o},x,y)$ in Eq. (\ref{tonec}).  This
is also clear since, if one considers the flows in the $p$-gons forming the
ends of the open chain, these have nowhere else to flow, and, since that is the
case, the same is true for the flows in all of the $p$-gons that form the
interior of the chain.

Using our calculation of the Tutte polynomial of $G_{e_1,e_2,e_g,m;c}$, 
we find, for the cyclic strip, 
\beq
F(G_{e_1,e_2,e_g,m;c},q) = \cases{ (q-1)(q-2)^m & if $e_g \ge 1$ \cr
                                   (q-1)(q-2)^m + (q-1)^m & if $e_g = 0$ } \ . 
\label{flow_nec}
\eeq
Thus, both for $e_g=0$ and for $e_g \ge 1$, the cyclic polygon chain graph
allows more $q$ flows than the open polygon chain graph.  This follows because
of the freedom of the flows in the cyclic case to make a global circuit around
the chain.  These calculations for the cyclic polygon chain may also be
compared with the result $F(C_n,q)=q-1$.  One sees that if $e_g=0$, then there
are more flows on the $G_{e_1,e_2,e_g,m;c}$ graphs, owing to the possibility of
flows within each polygon.  However, if $q=2$ and if $e_g \ge 1$, then, owing
to the degree-3 vertices where the polygons connect onto the line segments in
the chain, no $q$-flows can occur, since the flow conservation condition cannot
be satisfied at these vertices.

Since these results hold for arbitrary chain length $m$, one may consider the
limit of infinite length and define, as in Ref. \cite{f}, a function $\phi$
representing the number of $q$-flows per face of $G$ in this limit,
\beq
\phi(\{ G_{e_1,e_2,e_g;BC},q) = \lim_{m \to \infty} 
[F(G_{e_1,e_2,e_g,m;BC},q)]^{1/fc(G)} \ , 
\label{phi}
\eeq
where $fc(G)$ denotes the number of faces of $G$. Here, 
$fc(G_{e_1,e_2,e_g,m;o})=m+1$ and $fc(G_{e_1,e_2,e_g,m;c    })=m+2$.  Unlike the
free energy, the function $\phi$ does depend on whether one uses open or cyclic
boundary conditions, since some flows make global circuits around the chain in
the cyclic case.  For $q \ge 2$ (so that nowhere-zero $q$-flows can occur), we
calculate
\beq
\phi(\{G_{e_1,e_2,e_g;o}\},q) = \cases{ 
                 0 & if $e_g \ge 1$ \cr
               q-1 & if $e_g = 0$ } 
\label{phi_onec}
\eeq
and
\beq
\phi(\{G_{e_1,e_2,e_g;c}\},q) = \cases{ 
                 q-2 & if $e_g \ge 1$ \cr
                 q-1 & if $e_g = 0$ } \ . 
\label{phi_nec}
\eeq

\section{Reliability Polynomials}

A communication network, such as the internet, can be represented by a graph,
with the vertices of the graph representing the nodes of the network and the
edges of the graph representing the communication links between these nodes.
In analyzing the reliability of a network, one is interested in the probability
that there is a working communications route between any node and any other
node.  This is called the all-terminal reliability function.  This is commonly
modeled by a simplification in which one assumes that each node is operating
with probability $p_{node}$ and each link (abbreviated $\ell$) is operating
with probability $p_\ell$.  As probabilities, $p_{node}$ and $p_\ell$ lie in
the interval [0,1]. The dependence of the all-terminal reliability function
$R_{tot}(G,p_{node},p_\ell)$ on $p_{node}$ is an overall factor of
$(p_{node})^n$; i.e., $R_{tot}(G,p_{node},p_\ell) = (p_{node})^n R(G,p_\ell)$.
The difficult part of the calculation of $R_{tot}(G,p_{node},p_\ell)$ is thus
the part that depends on the links, $R(G,p_\ell)$.  The function 
$R(G,p_\ell)$ is given by 
\beq
R(G,p_\ell) = \sum_{\tilde G \subseteq G} p_\ell^{e(\tilde G)} \, 
(1-p_\ell)^{e(G)-e(\tilde G)}
\label{rgen}
\eeq
where $\tilde G$ is a connected spanning subgraph of $G$.  Each term in this
sum is the probability that the communication links $\tilde E \in \tilde G$ are
functioning (equal to $p_\ell^{e(\tilde G)}$) times the probability that the
other links, $E - \tilde E$, are not functioning (equal to
$(1-p_\ell)^{e(G)-e(\tilde G)}$).  From its definition, $R(G,p_\ell)$ is
clearly a monotonically increasing function of $p_\ell \in [0,1]$ with the
boundary values $R(G,0)=0$ and $R(G,1)=1$.  $R(G,p_\ell)$ is given in terms of
the Tutte polynomial, evaluated with $x=1$ (guaranteeing that $\tilde G$ is a
connected spanning subgraph of $G$) and $y=y_\ell$, where
\beq
y_\ell = \frac{1}{1-p_\ell} \quad i.e., \ v_\ell = y_\ell - 1 
= \frac{p_\ell}{1-p_\ell} \ , 
\label{yell}
\eeq
by the relation
\beq
R(G,p_\ell)=p_\ell^{n-1}(1-p_\ell)^{e(G)+1-n} \, T(G,1,\frac{1}{1-p_\ell}) \ . 
\label{rtrel}
\eeq

Using our calculation of the Tutte polynomials for 
$G_{e_1,e_2,e_g,m;o}$ and $G_{e_1,e_2,e_g,m;c}$, we find
\beq
R(G_{e_1,e_2,e_g,m;o},p_\ell) = \Big [ p_\ell^{p+e_g-1} 
\{ p(1-p_\ell) + p_\ell \} \Big ]^m
\label{ronec}
\eeq
and
\beqs
& & R(G_{e_1,e_2,e_g,m;c},p_\ell) = p_\ell^{(p+e_g-1)m} \Big [
  p(1-p_\ell)+p_\ell \Big ]^{m-1} \times \cr\cr
& \times & \Bigg [\bigg \{me_g(1-p_\ell)+p_\ell \bigg \} 
                  \bigg \{ p(1-p_\ell) + p_\ell \bigg \} \cr\cr
& + & \frac{m}{2}(1-p_\ell)^2 \bigg \{ p(p-1)-e_1(e_1-1)-e_2(e_2-1) \bigg \}
\Bigg ]  \ . 
\label{rnec}
\eeqs
In general, $R(G_{e_1,e_2,e_g,m;c},p_\ell) \ge R(G_{e_1,e_2,e_g,m;o},p_\ell)$,
with equality only at $p_\ell=0, \ 1$. This can be understood as a consequence
of the fact that with the cyclic boundary condition, there are more possible
communication routes linking two nodes than there are with the open boundary
condition.  We observe that for $p_\ell \in (0,1)$,
$R(G_{e_1,e_2,e_g,m;o},p_\ell)$ and $R(G_{e_1,e_2,e_g,m;c},p_\ell)$ are (i)
decreasing functions of $m$ for fixed $e_1$, $e_2$, and $e_g$; (ii) decreasing
functions of $e_g$ for fixed $e_1$, $e_2$, and $m$; and (iii) decreasing
functions of $e_1$ for fixed $e_2$, $e_g$, and $m$.  These properties can be
ascribed to the greater probability of communication bottlenecks as the
respective parameter, $m$, $e_g$, or $e_1$ increases with the other parameters
held fixed.  We also observe that for fixed $p=e_1+e_2$, $e_g$, and $m$, these
reliability polynomials increase as $|e_1-e_2|$ decreases.

As in \cite{r}, in the limit of infinite chain length, $m \to \infty$, one may
define a function $r$ that measures the reliability per node, as
\beq
r(\{G_{e_1,e_2,e_g}\},p_\ell) = \lim_{m \to \infty} [R(G_{e_1,e_2,e_g,m;BC},
p_\ell)]^{1/n} \ . 
\label{rasymp}
\eeq
As was discussed in \cite{r} for other strip graphs, this function is
independent of the longitudinal boundary conditions, so we drop the $BC$
subscript on the left-hand side of Eq. (\ref{rasymp}).  Clearly, for a general
$\{G\}$, $r(\{G\},p_\ell)$ is an increasing function of $p_\ell \in [0,1]$ with
the values $r(\{G\},0)=0$ and $r(\{G\},1)=1$. From our exact calculations
above, we find 
\beq
r(\{G_{e_1,e_2,e_g}\},p_\ell) = p_\ell \Big [p(1-p_\ell) + p_\ell 
\Big ]^{\frac{1}{p+e_g-1}} \ . 
\label{rasymp_chain}
\eeq
As a comparison, for the infinite-length limit of a line graph $L_n$ or circuit
graph $C_n$, $r(\{ {\cal L} \},p_\ell)=r(\{ {\cal C} \},p_\ell)=p_\ell \equiv
r_{_{1D}}(p_\ell)$.  Now for $p_\ell \in (0,1)$ (and $p \ge 2$), the factor
$[p(1-p_\ell)+p_\ell]^{1/(p+e_g-1)} > 1$, and hence
$r(\{G_{e_1,e_2,e_g}\},p_\ell) > r_{_{1D}}(p_\ell)$. Concerning the dependence
on $e_1$ and $e_2$ (which only enter in the form of their sum, $p$) and on
$e_g$, we find from an analysis of the respective partial derivatives, for a
fixed $p_\ell \in (0,1)$, that $r(\{G_{e_1,e_2,e_g}\},p_\ell)$ is a
monotonically decreasing function (i) of $p$, for fixed $e_g$, and (ii) of
$e_g$, for fixed $p$.  Furthermore,
\beq
\lim_{p \to \infty} r(\{G_{e_1,e_2,e_g}\},p_\ell) = p_\ell
\label{rplimit}
\eeq
and
\beq
\lim_{e_g \to \infty} r(\{G_{e_1,e_2,e_g}\},p_\ell) = p_\ell \ . 
\label{reglimit}
\eeq
Thus, in both of these limits, $r(\{G_{e_1,e_2,e_g}\},p_\ell)$ reduces to
$r_{_{1D}}(p_\ell)$.

\section{Percolation Clusters}

In this section we use our results to calculate a quantity of interest in the
area of bond percolation.  We first briefly mention some necessary background.
Consider a connected graph $G$ and assume that the vertices are definitely
present, but each edge is present only with a probability $p_\ell \in [0,1]$.
In the usual statistical mechanical context, one usually considers a limit in
which the number of vertices $n \to \infty$. An important quantity is the
average number of connected components (= clusters) in $G$.  For a given $G$,
we denote this average cluster number per vertex as $\langle k \rangle_G$. This
is given by
\beqs
\langle k \rangle_G & = & \frac{(1/n)\sum_{G^\prime} k(G^\prime)
p_\ell^{e(G^\prime)}(1-p_\ell)^{e(G)-e(G^\prime)} }{
\sum_{G^\prime} p_\ell^{e(G^\prime)}(1-p_\ell)^{e(G)-e(G^\prime)} } \cr\cr
& & 
\cr
& = &
\frac{(1/n)\sum_{G^\prime} k(G^\prime)v_\ell^{e(G^\prime)}}
{\sum_{G^\prime} v_\ell^{e(G^\prime)}} \ ,
\label{keq}
\eeqs
where $G^\prime$ is a spanning subgraph of $G$, as above, and $v_\ell$ was
defined in Eq. (\ref{yell}).  Hence, in the $n \to \infty$ limit, the average
cluster number per vertex, $\langle k \rangle_{\{G\}}$, is given by
\beq
\langle k \rangle_{\{G\}} = \frac{\partial f(\{ G \},q,v)}{\partial q} 
\bigg |_{q=1, \  v=v_\ell} \ . 
\label{kdfdq}
\eeq
Using Eq. (\ref{kdfdq}) with Eq. (\ref{fonec}), we find, for 
the infinite-length limits of both the open and cyclic polygon chains, the
average cluster number 
\beqs
\langle k \rangle_{\{G_{e_1,e_2,e_g}\}} & = & \frac{1}{(p+e_g-1)} \,
\bigg [ (p+e_g)(1-p_\ell)+p_\ell^p-1 \bigg ] \cr\cr
& = & \bigg (\frac{1-p_\ell}{p+e_g-1} \bigg ) \bigg [ p+e_g-
\sum_{j=0}^{p-1} p_\ell^j \bigg ] \ . 
\label{kave}
\eeqs
This may be compared with the result $\langle k \rangle_{1D}=1-p_\ell$ for the
infinite line.  We have
\beqs
\langle k \rangle_{1D} - \langle k \rangle_{\{G_{e_1,e_2,e_g}\}} & = & 
\bigg ( \frac{1-p_\ell}{p+e_g-1} \bigg ) \bigg [ (\sum_{j=0}^{p-1} p_\ell^j) -
  1 \bigg ] \cr\cr
& \ge & 0  \ , 
\label{kkdif}
\eeqs
with equality holding only if $p_\ell=1$ or $p=1$.  Thus, if $p_\ell < 1$ and
$p \ge 2$, the average number of clusters per vertex is greater for the line
than for the infinite-length limit of the polygon chain graph (with either set
of boundary conditions).

\section{Some Graphical Quantities}

Special valuations of the Tutte polynomial of a graph yield various quantities
describing properties of this graph.  In this section we give these. First, we
recall some definitions.  A tree graph is a connected graph with no circuits.
A spanning tree of a graph $G$ is a spanning subgraph of $G$ that is also a
tree.  A spanning forest of a graph $G$ is a spanning subgraph of $G$ that may
consist of more than one connected component but contains no circuits. The
special valuations of interest here are (i) $T(G,1,1)=N_{ST}(G)$, the number of
spanning trees ($ST$) of $G$; (ii) $T(G,2,1)=N_{SF}(G)$ the number of spanning
forests ($SF$) of $G$; (iii) $T(G,1,2)=N_{CSSG}(G)$, the number of connected
spanning subgraphs ($CSSG$) of $G$; and (iv) $T(G,2,2)=N_{SSG}(G)=2^{e(G)}$,
the number of spanning subgraphs ($SSG$) of $G$.  For both the open and cyclic
strips, the last of these quantities is directly determined by Eq. (\ref{enec})
(without the necessity of calculating the Tutte polynomial) to be
\beq
N_{SSG}(G_{e_1,e_2,e_g,m;o})=N_{SSG}(G_{e_1,e_2,e_g,m;c}) = 2^{(p+e_g)m} \ . 
\label{ssg_enec}
\eeq
We evaluate our general results for $T(G_{e_1,e_2,e_g,m;o},x,y)$ and 
$T(G_{e_1,e_2,e_g,m;c},x,y)$ to obtain the quantities (i)-(iii). 
For the numbers of spanning trees, we find 
\beq
N_{ST}(G_{e_1,e_2,e_g,m;o}) = p^m
\label{st_onec}
\eeq
and
\beq
N_{ST}(G_{e_1,e_2,e_g,m;c}) = mp^{m-1}\bigg [ p e_g + \frac{1}{2} \Big \{
p(p-1)- e_1(e_1-1) - e_2(e_2-1) \Big \} \bigg ] \ . 
\label{st_nec}
\eeq
The numbers of spanning forests are
\beq
N_{SF}(G_{e_1,e_2,e_g,m;o}) = \bigg [ 2^{e_g}(2^p-1) \bigg ]^m 
\label{sf_onec}
\eeq
and
\beq
N_{SF}(G_{e_1,e_2,e_g,m;c}) = \bigg  [ 2^{e_g}(2^p-1) \bigg ]^m 
- \bigg [ 2^{e_1}+2^{e_2}-2 \bigg ]^m \ . 
\label{sf_nec}
\eeq
The numbers of connected spanning subgraphs are
\beq
N_{CSSG}(G_{e_1,e_2,e_g,m;o}) = (p+1)^m 
\label{cssg_onec}
\eeq
and
\beqs
N_{CSSG}(G_{e_1,e_2,e_g,m;c}) & = & (p+1)^{m-1} \bigg [ (me_g+1)(p+1) \cr\cr
& + & \frac{m}{2} \Big \{ p(p-1) - e_1(e_1-1) - e_2(e_2-1) \Big \} \bigg ] \ .
\label{cssg_nec}
\eeqs

We also give another evaluation of the Tutte polynomial. For a general
connected graph $G=(E,V)$, one can define an orientation of of $G$, i.e., a
directed graph ${\vec G}=(V,{\vec E})$ by assigning a direction to each edge $e
\in E$.  There are $2^{e(G)}$ of these orientations. Among these, an acyclic
orientation of $G$ is defined as an orientation that does not contain any
directed cycles. Here, a directed cycle is a cycle in which, as one travels
along the cycle, all of the oriented edges have the same direction.  The number
of such acyclic orientations is denoted $a(G)$ and is given by the evaluation
of the Tutte polynomial with $x=2$ and $y=0$ \cite{stanley}:
\beq
a(G)= T(G,2,0) \ . 
\label{at}
\eeq
Equivalently, this is obtained by the evaluation of the chromatic polynomial
at $q=-1$: $a(G)= (-1)^{n(G)}P(G,-1)$.  From the results in \cite{nec}, in
agreement with our calculations here, we have, for the numbers of acyclic
orientations of the open and cyclic polygon chain graphs 
\beq
a(G_{e_1,e_2,e_g,m;o}) = \Big [2^{e_g}(2^p-2) \Big ]^m
\label{aonec}
\eeq
and
\beq
a(G_{e_1,e_2,e_g,m;c}) = \Big [2^{e_g}(2^p-2) \Big ]^m 
-2\Big [ 2^{e_1}+2^{e_2}-3 \Big ]^m \ . 
\label{anec}
\eeq

\section{Conclusions}

In conclusion, in this paper, generalizing our previous results on chromatic
polynomials with S.-H. Tsai in \cite{nec}, we have presented exact calculations
of the Potts model partition functions and equivalent Tutte polynomials for a
class of polygon chain graphs with open and cyclic boundary conditions.  We
have evaluated special cases of these results to compute the corresponding flow
polynomials, reliability polynomials, and various quantities of graph-theoretic
interest, and have analyzed the dependence on the parameters $e_1$, $e_2$, and
$e_g$ characterizing the families of graphs. 

\section{Acknowledgments}

This research was partially supported by the grant NSF-PHY-06-53342.

\end{document}